\documentclass{article}
\usepackage[utf8]{inputenc}
\usepackage[english]{babel}

\usepackage[final]{style} 
\usepackage{hyperref}
\usepackage[table]{xcolor} 
\usepackage{graphicx}
\usepackage{textcomp}
\usepackage{wrapfig}
\usepackage{multirow}
\usepackage{amsmath,amssymb,amsfonts}
\usepackage{amsthm}
\usepackage{mathrsfs}
\usepackage{manyfoot}
\usepackage{booktabs}
\usepackage{algpseudocode}
\usepackage{listings}
\usepackage{subcaption}  

\usepackage{natbib}
\setcitestyle{square,numbers,compress,sort}

\def\model{FluxBrain}
 
\title{Looking through the mind's eye via multimodal encoder-decoder networks}
 
\author{Arman Afrasiyabi$^{\star \diamond \bullet}$, 
Erica Busch$^{\dag \ast \bullet}$, \\
\textbf{Rahul Singh$^{\star \ast \bullet}$, Dhananjay Bhaskar$^{\star \diamond \bullet}$, Laurent Caplette$^{\dag \ast \bullet}$}, 
\\ \textbf{Nicholas Turk-Browne$^{\dag \ast \bullet}$, Smita Krishnaswamy$^{\star \diamond \circ \ast \bullet}$} \\
$^{\star}$ Department of Computer Science,  
$^{\circ}$ Applied Mathematics Program \\
$^{\dag}$ Department of Psychology, 
$^{\diamond}$ Department of Genetics, \\
$^{\ast}$ Wu Tsai Institute,
$^{\bullet}$ Yale University
}

\newcommand{\unnumberedfootnote}[1]{%
  \let\thefootnote\relax\footnotetext{#1}%
  \let\thefootnote\svthefootnote%
}

\begin{document}

\maketitle

\begin{abstract}
\unnumberedfootnote{\color{blue} This work is accepted and presented at the Center for Collaborative Arts and Media, Yale University.}
In this work, we explore the decoding of mental imagery from subjects using their fMRI measurements. In order to achieve this decoding, we first created a mapping between a subject's fMRI signals elicited by the videos the subjects watched. This mapping associates the high dimensional fMRI activation states with visual imagery. Next, we prompted the subjects textually, primarily with emotion labels which had no direct reference to visual objects. Then to decode visual imagery that may have been in a person's mind's eye, we align a latent representation of these fMRI measurements with a corresponding video-fMRI based on textual labels given to the videos themselves. This alignment has the effect of overlapping the video fMRI embedding with the text-prompted fMRI embedding, thus allowing us to use our fMRI-to-video mapping to decode.  Additionally, we enhance an existing fMRI dataset, initially consisting of data from five subjects, by including recordings from three more subjects gathered by our team. We demonstrate the efficacy of our model on this augmented dataset both in accurately creating a mapping, as well as in plausibly decoding mental imagery. 
\end{abstract}

\section{Introduction}

Humans have always endeavored to bring their thoughts and imaginations into tangible reality. In the 20th and 21th centuries, movements like Impressionism and Surrealism revolutionized the art world by depicting the stream of consciousness and dream-like states, pushing the boundaries of visual representation. In recent years, the field of AI-generated art has made significant strides, with algorithms like DALL$\cdot$E~\cite{ramesh2022hierarchical}, Midjourney~\cite{oppenlaender2022creativity}, and Stable Diffusion~\cite{ho2020denoising}, creating intricate images based on textual prompts provided by humans, further bridging the gap between thought and visual expression. Now, advances in neuroimaging and artificial intelligence are poised to unlock unprecedented capabilities, allowing for the direct decoding of images from human thought, thereby offering a new frontier in the visualization of the mind's eye.

The trajectory of functional magnetic resonance imaging (fMRI)-based visual information reconstruction has rapidly evolved, initially relying on traditional decoding techniques~\cite{kamitani2005decoding, haynes2005predicting, cox2003functional, haxby2001distributed} and advancing towards hyperrealistic reconstructions with the integration of deep learning and GANs~\cite{goodfellow2014generative, haxby2001distributed, dado2022hyperrealistic, shen2019deep, seeliger2018generative, ozcelik2022reconstruction, lin2022mind}. The introduction of diffusion~\cite{ho2020denoising} and stable diffusion models~\cite{rombach2022high} has further refined fMRI-to-image reconstructions, achieving unparalleled quality and demonstrating the potential of neural imaging technology in decoding and visualizing human cognition~\cite{chen2023seeing, takagi2023high, ozcelik2023brain, gu2022decoding}. This evolution underscores the merging of generative artificial intelligence with neuroimaging data, leading to breakthroughs in the direct translation of neural activity into visual representations and offering profound implications for aiding individuals with disabilities~\cite{scotti2024mindeye2,scotti2024reconstructing,ozcelik2022reconstruction,ozcelik2023brain,gu2022decoding}.

The existing methods are primarily trained on neuroimaging datasets obtained during image viewing to recreate those images, however, they fall short of fully capturing the nuanced spectrum of human thought and imagery. This is evident when considering the visualization of images that participants spontaneously generate in their minds. The core challenge lies in narrowing the gap between the tangible data used for training such models and the abstract, multifaceted nature of human thoughts and imagination. Our methodology aims to mitigate this limitation by aligning the representations of neuroimaging data with those of videos and imagined scenes. This approach transcends the mere replication of observed visuals; it endeavors to decode the thoughts evoked by a piece of text, identifying the image representations that best resonate with the neuroimaging activity elicited by our thoughts. This effort is about more than capturing snapshots of brain activity; it's about deciphering the underlying neural mechanisms and converting them into a visual language understandable by AI. In doing this, we can reproduce not only the images a subject directly observes but also create visual representations of abstract thoughts and concepts, offering a fuller understanding of the cognitive and perceptual processes involved.

In this work, we introduce a novel method that captures and visually represents individuals' thoughts stimulated textually, advancing beyond the conventional technique of reconstructing images from direct brain recordings triggered by visual stimuli. Our approach employs an innovative encoder-decoder-based algorithm, meticulously designed to link the visual representations encountered in scenes—such as those found in videos and images—with the brain recordings obtained during these visual experiences. This link is established during our model’s training phase. Subsequently, in the inference phase, our model innovatively generates and reconstructs visual appearances based on brain activity elicited by text prompts. Essentially, our algorithm translates the abstract thoughts captured in these recordings into vivid visual imagery, effectively aiming to "visualize" the dreams or thought processes of individuals based on their brain activity.

This paper presents the following contributions. First, we introduce a novel multi-encoder-decoder architecture designed for effective training through both point-to-point and distribution-to-distribution matching between brain recordings and video samples. Secondly, we present the brain decoding inference method with a pre-trained diffusion model aimed at enhancing the quality of the decoded brain states. Thirdly, we expand upon an existing fMRI dataset, originally comprising data from five subjects as detailed in~\cite{horikawa2020neural}. We enrich this dataset by incorporating recordings from three additional subjects, collected by our research group, thereby broadening the scope and applicability of the dataset. Finally, we showcase the effectiveness of our proposed model through its application to this enhanced dataset. The results demonstrate that our model can reconstruct brain visualization of text in a manner that is both conceptually meaningful and consistent over time, underscoring the potential of our approach in decoding and visualizing human brain states.

\section{Method}

\begin{figure}  
    \centering
    \begin{subfigure}[b]{0.285\textwidth}
        \includegraphics[width=\textwidth]{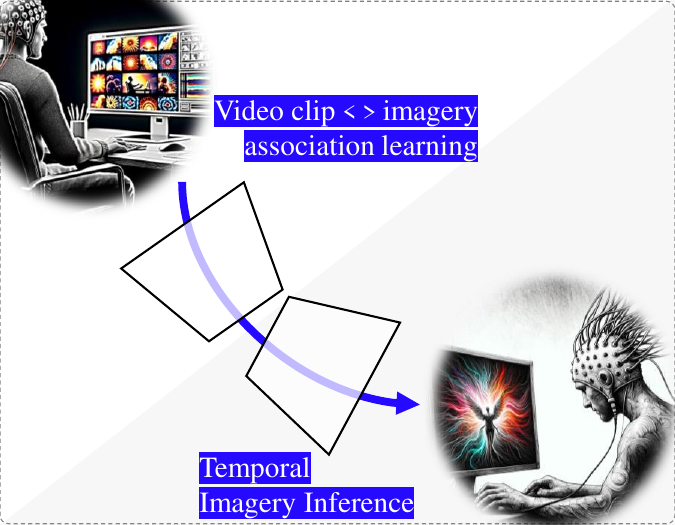}
        \caption{idea}
        \label{fig:idea}
    \end{subfigure}
    \begin{subfigure}[b]{0.40\textwidth}
        \includegraphics[width=\textwidth]{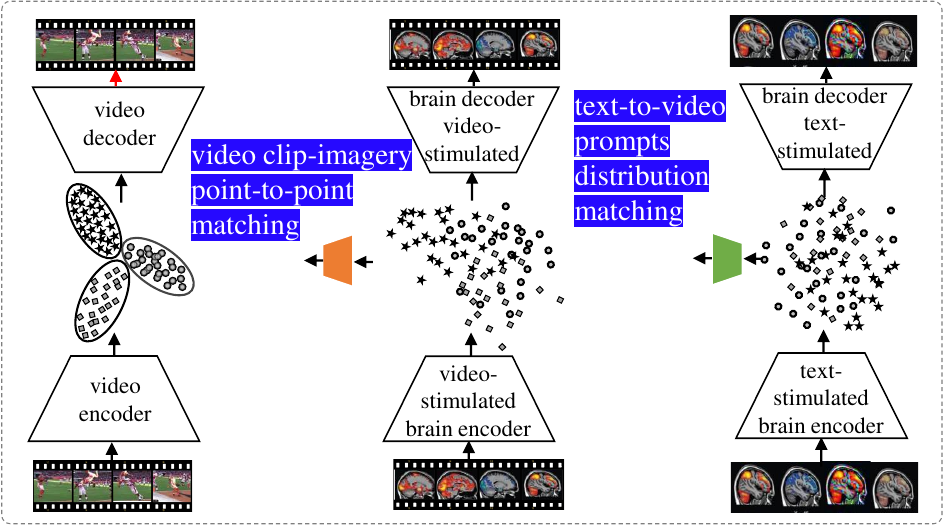}
        \caption{\model{}}
        \label{fig:model}
    \end{subfigure} 
    \begin{subfigure}[b]{0.19\textwidth}
        \includegraphics[width=\textwidth]{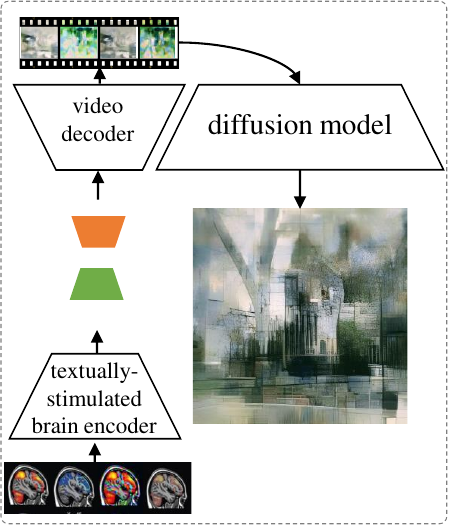}
        \caption{inference}
        \label{fig:diff_model}
    \end{subfigure} 
    \caption{In (a), our model provides a schematic overview, detailing its initial focus on learning the connections between video clips and imagery through machine learning techniques. Subsequently, it generates artistic visualizations based solely on brain recordings. (b) illustrates the architecture of our \model{}, comprising three distinct encoder-decoder models. These models are responsible for reconstructing video (left), brain activity during video stimulation (middle), and recovering text prompts (right). An orange network facilitates one-to-one matching between the embeddings of video and stimulus-brain recordings, while a green network aims at aligning the distributions of text- and video-stimulus prompts.  (c) presents the inference pathway leading to the reconstruction of brain activity stimulated by text, utilizing our proposed architectural framework.  
     }
    \label{fig:overview_and_x2y}
\end{figure}

The objective of this work is to establish a connection between cutting-edge generative models (images/videos, audio, and text) and human cognition or imagination represented via the corresponding brain recordings. In particular, we aim to comprehend how video clips can influence the model responsible for temporal imaginative inference, as illustrated in Figure~\ref{fig:idea}. To achieve this, we leverage an encoder-decoder model framework to introduce a method for text-based stimulation retrieval. This is facilitated by our newly developed guided generative models, which are adept at distribution matching. This approach significantly improves the accuracy of representation extraction from brain recordings, providing a novel pathway to understanding and enhancing the interface between artificial intelligence and human cognitive functions.

As depicted in Figure \ref{fig:model}, we employ three encoder-decoder models(to infer the representations of video, text, and fMRI) aimed at reconstructing visual representations from brain recordings triggered by textual stimuli. Our approach incorporates three generative models to facilitate the creation of videos or images, specifically focusing on video reconstruction, brain stimulation, and a textual-stimulated encoder-decoder framework. Notably, our video-simulated brain decoder is designed to reconstruct brain recordings (such as those obtained from fMRI scans) by first mapping these recordings onto the video embeddings derived from the video encoder. Furthermore, we introduce an additional mapping function, visually represented by the orange model in Figure \ref{fig:model}, which facilitates the interaction between the video and the video-stimulated encoder-decoders. This enables our model to learn associative alignments through point-to-point matching techniques. 
Mathematically, the initial encoder, denoted as $g_\text{v}(\cdot|\phi_\text{v})$, and decoder, denoted as $f_\text{v}(\cdot|\theta_\text{v})$, form a video-specific model architecture designed to learn representations by reconstructing video frames, thereby facilitating the learning of embeddings.
\begin{align} \nonumber
    \mathcal{L}_{\text{v}} = \frac{1}{|S_\text{v}|} \sum_{ \mathbf{x}_{\text{v}, i}\in S_\text{v}} ||\mathbf{x}_{\text{v}, i}- f_\text{v}(g_\text{v}(\mathbf{x}_{\text{v}, i}|\phi_\text{v}) | \theta_\text{v}) ||^2, 
\end{align}
where $\mathbf{x}_{\text{v}, i}$ is a frame of a video $S_\text{v}$. Beside, $\theta_\text{v}$ and $\phi_\text{v}$ are the parameters of the $f_\text{v}$ and $g_\text{v}$, respectively.

Equipped with the video reconstruction autoencoder, we proceed to construct an fMRI encoder, \(g_{\text{f}}(\cdot|\phi_{\text{f}})\), and decoder, \(f_{\text{f}}(\cdot|\theta_{\text{f}})\), architecture specifically tailored to learn the representations of brain recordings stimulated by video. With these dual encoder-decoder architectures for video and fMRI, we now build a bridge to create a learning module that translates the fMRI encoder's output into the video encoder's input. To this end, we employ a lightweight mapping function, \(\text{MAP}_{\text{f} \rightarrow \text{v}}(\cdot) = q(\cdot | W_{\text{q}})\), that aims at point-to-point matching fMRI embeddings with their corresponding video frames to which the brain has been exposed. Given these elements, the overall loss function for training the fMRI encoder-decoder and refining the encoder for both fMRI and video would be as follows:
\begin{align} \nonumber
    \mathcal{L}_{\text{f}} = \frac{1}{|S_\text{f}|} \sum_{ \mathbf{x}_{\text{f}, i}\in S_\text{f}} ||\mathbf{x}_{\text{f}, i}- f_\text{f}(g_\text{f}(\mathbf{x}_{\text{f}, i}|\phi_\text{f}) | \theta_\text{f}) ||^2    
    + \frac{1}{|S_\text{f}|} 
    \sum_{\substack{(\textbf{x}_{\text{v}, i}) \in S_\text{v} \\ (\textbf{x}_{\text{f}, i}) \in S_\text{f} }} \|  g_\text{v}(\mathbf{x}_{\text{v}, i}|\phi_\text{v}) -     
    \text{MAP}_{\text{f} \rightarrow \text{v}}(g_\text{f}(\mathbf{x}_{\text{f}, i}|\phi_\text{f})\|^2, 
\end{align}
where $\mathbf{x}_{\text{f}, i}$ is a frame of an fMRI recording while the brain has been stimulated by video $S_\text{f}$, and $|S_\text{f}| = |S_\text{v}|$. Beside, $\theta_\text{f}$ and $\phi_\text{f}$ are the parameters of the $f_\text{f}$ and $g_\text{f}$, respectively.

After configuring the video-stimulated brain encoder, we proceed by freezing the operations of the first and second networks while focusing on training the third network. This network is specifically designed to align with text-based stimuli, such as textual content as presented in Figure~\ref{fig:model}. However, given the complexity of creating a direct point-to-point alignment in constructing a textually stimulated brain model, we shift our approach to analyzing the overall distribution patterns of textual-stimulated brain recordings. These patterns are compared against the embedding space generated by the video-stimulated architecture 
To achieve this, our next step is to train a text-prompted fMRI encoder, \(g_{\text{e}}(\cdot|\theta_{\text{e}})\), and decoder, \(f_{\text{e}}(\cdot|\phi_{\text{e}})\).  
Thus, we aim to align the text-based distribution with that of the distribution derived from visually and textually stimulated brain recordings. Despite science videos or visual content possessing textual labels, we undertake the task of text-based distribution-to-distribution matching. Our strategy involves guiding the embeddings of textually stimulated brain recordings closer to those from visually stimulated recordings. Motivated by prototypical networks~\cite{snell2017prototypical}, we begin by calculating the prototype for the \(k\)-th class \(C\) in the visually stimulated fMRI context, denoted by \(\mathbf{p}_\text{v}^k\):
\begin{align} \nonumber
    \mathbf{p}_\text{f}^k &= \frac{1}{|S^k_\text{f}|} \sum_{(\mathbf{x}_\text{f}^i, {y}^i) \in S^k_\text{f}} g_{\text{f}}(\mathbf{x}_\text{f}^i | \phi_\text{f}),
\end{align}
where \(\mathbf{p}_\text{f}^k\) represents the centroid of the fMRI recordings for subjects watching a video from the \(k\)-th class. We subsequently apply an MLP-based mapping function, \(\text{MAP}_{\text{e} \rightarrow \text{f}}(\cdot) = h(\cdot|W_{\text{h}})\), to align textually stimulated fMRI recordings with visually stimulated fMRI recordings that share the same class label. To achieve this, we use a distance function \(d(\cdot, \cdot)\). Specifically, we define a cross-entropy loss function to perform textual-based distribution matching based on the following softmax function:
\begin{align}
    P_{\phi} (y=k | \mathbf{p}_\text{f}^k) = 
    \frac{\exp\big(-d \big(\text{MAP}_{\text{e} \rightarrow \text{f}}(g_{\text{e}}(\mathbf{x}^i_{\text{e}}|\phi_\text{e})), \mathbf{p}_\text{f}^k\big)\big)}
    {\sum_{k'}\exp\big(-d \big(\text{MAP}_{\text{e} \rightarrow \text{f}}(g_{\text{e}}(\mathbf{x}^i_{\text{e}}|\phi_\text{e})), \mathbf{p}_\text{f}^{k'}\big)\big)}, 
\end{align}
Moreover, we employed a mean-squared loss function for representation learning of the textually stimulated brain recordings by both the encoder and decoder. The learning process progresses by minimizing the MSE loss function and or negative log-probability for distribution matching \(\mathcal{L}_{\text{matching}} = -\log P_{\phi}(y=k | \mathbf{x})\) of the correct class \(k\) through Stochastic Gradient Descent (SGD). 
\begin{align}
    \mathcal{L}_{\text{e}} = \frac{1}{|S_\text{e}|} \sum_{ \mathbf{x}_{\text{e}, t}\in S_\text{e}} ||\mathbf{x}_{\text{e}, t}- f_\text{e}(g_\text{e}(\mathbf{x}_{\text{e}, t}|\phi_\text{e}) | \theta_\text{e}) ||^2 + \mathcal{L}_{\text{matching}},
\end{align}
where $\mathbf{x}_{\text{e}, t}$ is the brain recording captured at time $t$. Beside, $\theta_\text{e}$ and $\phi_\text{e}$ are the parameters of the $f_\text{e}$ and $g_\text{e}$, respectively. 

Furthermore, as illustrated in Figure \ref{fig:diff_model}, we improve the quality of reconstructed images through the application of an advanced large diffusion generative model. This method involves a sophisticated process where the diffusion model iteratively refines the image, gradually enhancing its clarity, detail, and overall fidelity to the original text or cognitive stimulus. Specifically, this approach starts by generating a basic outline or low-resolution version of the target image based on the initial brain recordings. The large diffusion model then incrementally applies a series of transformations, each designed to enhance the image's accuracy and detail, by closely aligning it with the associated textual or cognitive data. This process leverages the power of deep learning to simulate and amplify the intricate process of human perception and memory reconstruction and produce high-quality images that more accurately reflect the subtleties and complexities of the original brain recordings. 
\section{Dataset}
 
In this project, we use an open-source dataset of short video clips, validated to be evocative of a wide range of emotions ~\cite{cowen2017self}. The dataset contains 2,185 videos (0.15 -- 90s) and across two experiments, participants provided ratings of the videos' emotional contents, uncovering 27 dimensions of emotion categories. In a separate study, researchers presented the same short videos to five participants (1 female, mean age = 25.8 years) undergoing fMRI scans. Participants were instructed to simply pay attention to the videos. Videos were edited to an 8s presentation time, with a 2s fixation cross between presentations. This resulted in approximately 8 hours of fMRI scanning per participant~\cite{horikawa2020neural}.

In our experiments, we extended beyond the original study by Horikawa and colleagues~\cite{horikawa2020neural}. First, we identified emotions of interest from the video dataset released by Cowen and colleagues~\cite{cowen2017self}, and sorted the videos according to their primary emotion labels. Using the matched fMRI dataset, we identified stimuli that resulted in the most ``emotionally charged" brain activation patterns, across participants. In other words, the fMRI responses to these stimuli were reliably predictive of the primary emotion label identified. As in the study by Horikawa et al.~\cite{horikawa2020neural}, we presented these stimuli to three participants (1 female, mean age = 29.3 years) in the fMRI scanner, looping the videos for 8s with a 2s rest period between. However, we interspersed the video trials with emotion-matched imagery trials. During these trials, participants were presented with a text prompt of a single emotion category from the emotions of interest we identified (i.e., ``joy'', ``surprise''), which stayed on the screen for the 8s duration. Participants were instructed to imagine ``in their mind's eye" a memory or an image that evokes that emotion; for instance, they could envision a birthday party from their childhood for ``joy''. Categories were repeated each 10 times across the trials, and participants were instructed to conjure the same image each time.

\section{Evaluation}

\paragraph{Implementation detail}
Our proposed methodology was developed using the PyTorch library, leveraging an NVIDIA A100 GPU with 80GB of VRAM for computational power. To train our network effectively in each iteration, we selected a random batch of sample sets categorized by emotion labels. These batches encompassed three distinct types of data: frames from video clips, fMRI recordings triggered by video stimuli, and fMRI recordings stimulated by emotional text. For the video processing component, we crafted a four-layer UNet architecture, which is essentially a 2D convolutional neural network (CNN) designed for encoding and decoding video information. 
\begin{wrapfigure}[15]{r}{0.30\textwidth}
    \vspace{-5mm}
    \centering
    \includegraphics[width=0.30\textwidth]{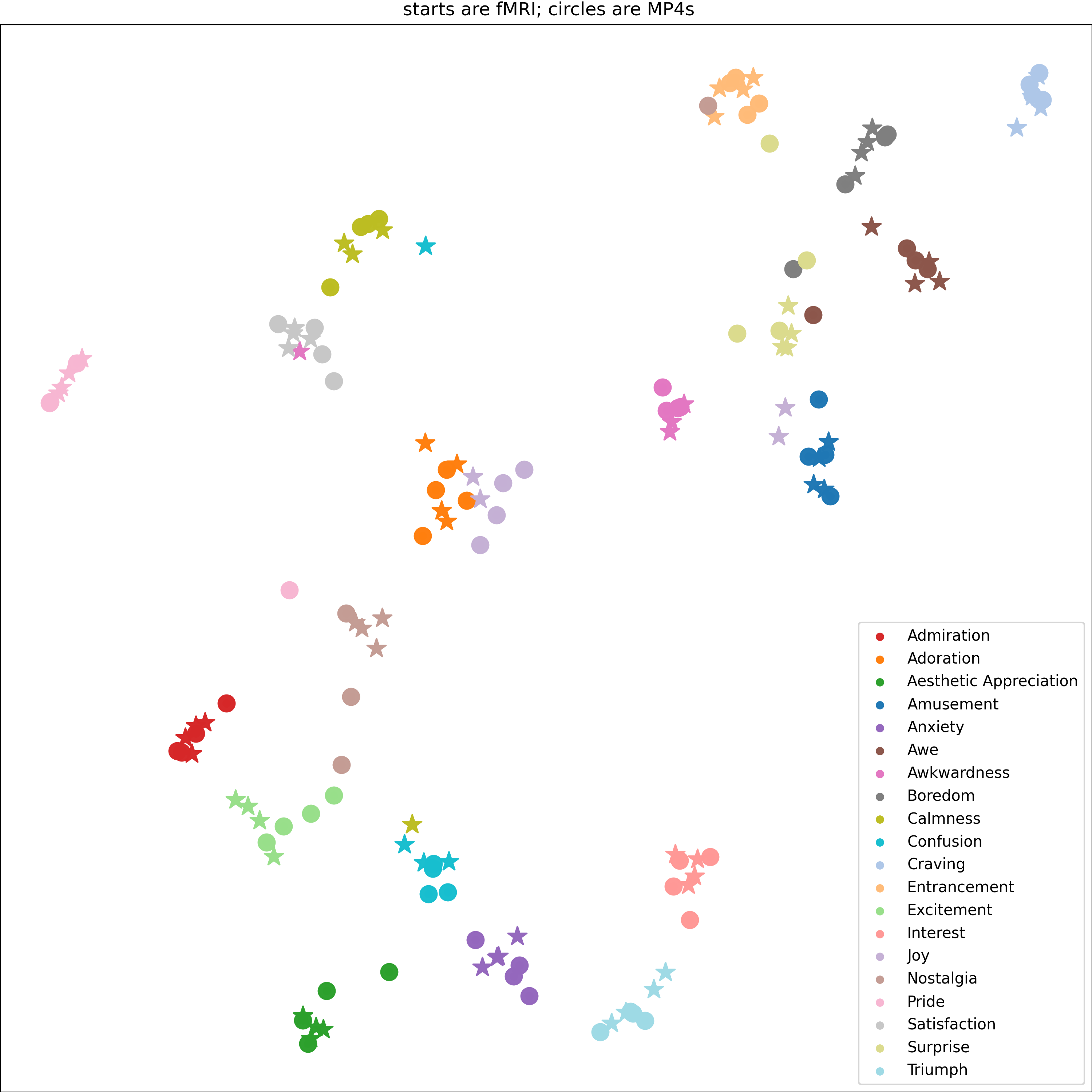}
    \caption{\footnotesize PHATE~\cite{moon2019visualizing} visualization showcasing the distribution alignment between video prompts and their corresponding emotion expressed as text.}
     \label{fig: phate_visualization}
\end{wrapfigure}
Additionally, for the fMRI data, we designed an autoencoder based on a 1D CNN. The training process utilized the Adam optimization algorithm, with a set learning rate of 0.0005, to adjust the network's parameters efficiently.

\paragraph{Distribution matching}

Figure~\ref{fig: phate_visualization} displays a PHATE~\cite{moon2019visualizing} visualization, highlighting the embedding space after training our model and performing inference. In this visualization, emotions are differentiated by color coding, with fMRI recordings represented as stars and video embeddings depicted as circles. The figure demonstrates our model's capability in achieving distribution matching between fMRI data and video frames within the embedding space, effectively bridging the gap between neural recordings and visual stimuli.

\begin{figure}
    \centering
    \includegraphics[width=0.99\linewidth]{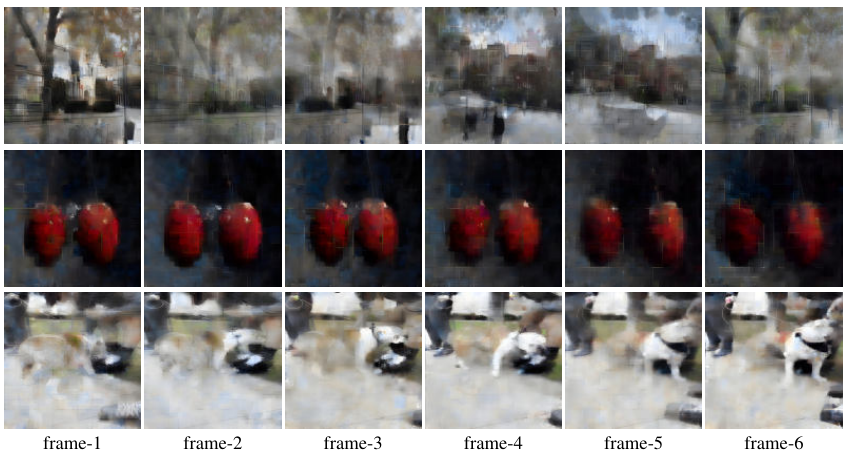}
    \caption{Video-prompt fMRI reconstruction based on brain recordings stimulated by video prompts, with each row representing a unique video simulation scenario. }
    \label{fig: results_video_reconstruction} 
\end{figure}

\paragraph{Reconstruction}
Figure~\ref{fig: results_video_reconstruction} showcases our model's capability in reconstructing videos from fMRI recordings, using video prompts such as the Yale-Sterling Library, Motion Cherry, and Yale's Kingman. These examples, excluded from the training set, demonstrate the model's proficiency in accurately capturing both the location and motion dynamics. As depicted in Figure~\ref{fig: results_video_reconstruction}, the reconstructed visuals affirm the model's effectiveness, offering clear and visually appealing representations of the original stimuli. 

\begin{figure}
    \centering
    \includegraphics[width=0.99\linewidth]{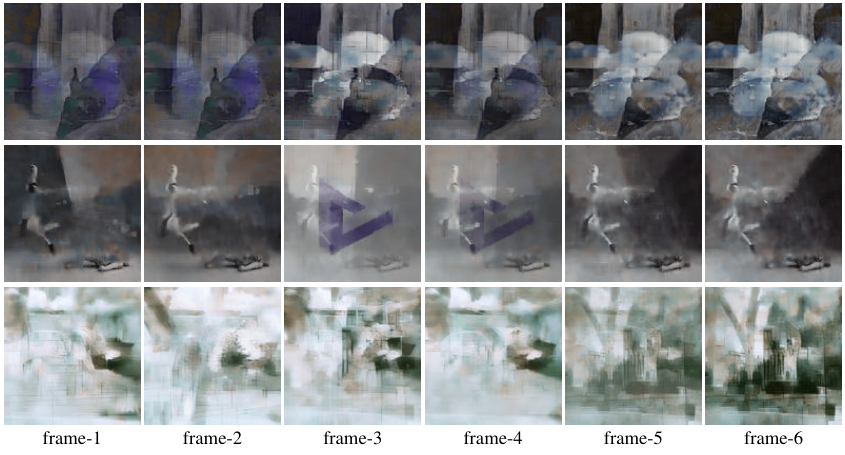}
    \caption{Text-prompt fMRI reconstruction from brain recordings triggered by text prompts, where each row displays a distinct emotion simulation scenario across six sequential frames.}
    \label{fig: results_fmri_reconstruction} 
\end{figure}

Figure~\ref{fig: results_fmri_reconstruction} demonstrates our model's ability to reconstruct the overall essence of emotions and text prompts into visual form. Specifically, it showcases the reconstruction of emotions associated with "Carving" in the first row, "Entrancement" in the second row, and "Yale" in the third row. The figure illustrates that our model achieves conceptually coherent results. For instance, the last frame related to the Yale prompt reveals a scene with a green landscape and a building, suggesting a particular location reminiscent of what the subject envisioned. These results stem from test sets that were excluded from the training phase. According to subjects' reports, the reconstructions conceptually align with their imagined scenarios during data collection.

\section{Discussion}

This work is a step forward towards visualization of the mind’s eye, i.e., decoding of imagery from human thought. The proposed technology holds the promise to be employed in multiple fronts including aiding individuals with disabilities and diagnosis. For individuals with communication impairments, such as those resulting from strokes or traumatic brain injury, this technology can restore the ability to communicate through the reconstruction of their visual thoughts into images or text, thereby providing a new form of expression for those who are unable to speak or write. Moreover, the ability to decode images and video from brain activity could revolutionize the field of neurology and psychiatry by providing unprecedented insights into brain function. It could aid in the diagnosis of neurological disorders, such as Alzheimer's, epilepsy, or brain injuries, by allowing doctors to observe the visual and sensory experiences of patients directly. Additionally, it could improve the understanding and treatment of mental health conditions like PTSD, schizophrenia, and depression. However, in addition to the promises of this technology of visualizing the mind’s eye, one must also be aware of the following implications.

\paragraph{Privacy and Ethical Implications} The most immediate concern is the invasion of privacy. With machine learning models being able to decode and reconstruct visual experiences from brain activity, they could potentially access the most private and intimate thoughts of an individual without their consent. Ensuring that such technology is used only with explicit, informed consent is crucial. Moreover, the purpose for which this technology is used must be ethically justified. While it could have beneficial applications, such as helping individuals who cannot communicate verbally to express themselves, it could also be used for nefarious purposes, such as surveillance or manipulating thoughts and behaviors.

\paragraph{Security, Accuracy and Interpretation} The risk of unauthorized access or hacking into this sensitive data could lead to unprecedented levels of personal exposure and blackmail. Strong security measures would be essential to protect individuals from such vulnerabilities. The reliability of the technology is another concern. Misinterpretation of the decoded images or videos could lead to false conclusions about an individual's thoughts or intentions, with potentially severe consequences for personal freedom and privacy.

\paragraph{Psychological Impact and Regulatory Framework} Knowing that one's visual thoughts could be decoded and viewed by others could lead to psychological stress and a sense of lost autonomy over one's inner life. The impact on mental health and the concept of personal identity would need careful consideration. Moreover, the development and use of this technology would necessitate robust legal frameworks to regulate its use, protect individuals' rights, and ensure that ethical standards are maintained. These frameworks would need to balance the potential benefits of the technology with the risks to individual rights and freedoms.

\section{Acknowledgments}
We would like to express our thanks to E. Chandra Fincke for assistance with data collection. We also acknowledge the use of DALL$\cdot$E for the generating of the graphic illustration presented in Figure 1(a), both were created based on our designed prompts.  

\bibliography{egbib}

\end{document}